\documentclass[aps,prl,floatfix,twocolumn]{revtex4-1}

\usepackage{graphicx}
\usepackage{braket}
\def \ingan {In$_x$Ga$_{(1-x)}$N~}
\begin{document}
\title{Observation of lattice thermal waves by the Blinking in photoluminescence of InGaN Quantum Well devices}

%% Notice placement of commas and superscripts and use of &
%% in the author list

\author{R. Micheletto}\author{K. Oikawa}\author{C. Feldmeier}

\affiliation{ Nanoscience and Technology, International Graduate School of Art and Sciences, Yokohama City University, Seto, Kanazawa-ku, Yokohama 236-0027, Japan}
%\author{U. T. Schwarz}
%\affiliation{Fraunhofer Institute for Applied Solid State Physics (IAF), Tullastrasse 72, 79108 Freiburg, Germany}

\begin{abstract}
The photoluminescence of III-V wide band-gap semiconductors as InGaN is characterized by local intensity fluctuations, known as 'blinking points', that despite decades of research are not yet completely understood. In this letter we report experimental data and a theoretical interpretation that suggests they are caused by the interference of thermal vibrations of the Quantum Well lattice. With far-field optical tests we could observe the lower frequency tail of these interference waves and study their dynamics as they propagate up to distances of several tens of microns.
\end{abstract}

\maketitle

\section{Introduction}
The phenomenon called 'photoluminescence blinking' have been observed in confined structures, as for example in semiconductors nanocrystals \cite{nirm:1996}. Because of the local nature of these systems, the theoretical treatment can be zero-dimensional (considering only time-energy dependence in one point). The blinking process results to be related to temporary quenching of photoluminescence due to highly efficient non-radiative recombination processes as for example Auger effect or other non-radiative processes\cite{bross:2010}. Nevertheless, in InGaN devices the band structure is an infinitely wide quantum well, so a zero-dimensional model is insufficient, the understanding of the blinking phenomenon remains elusive \cite{mich:2009,stef:2009,mich:2006a,kawa:2003}. \par

\begin{figure}[!h]
\begin{center}\leavevmode
\includegraphics[width=8.5cm]{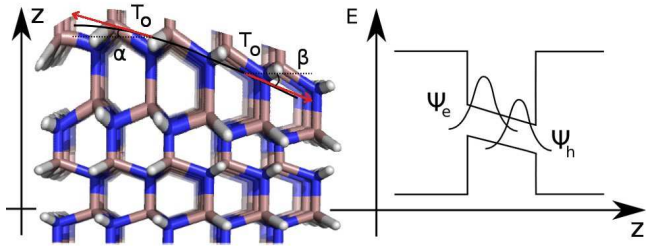}
\caption{An exaggerated scheme in which the Wurtzite structure is stretched along the crystal growth direction Z (left) and the corresponding Energy density diagram (right). The mechanical stress along the line of adjacent atoms indicated by the line, is represented by the tension $T_o$ shown in two arbitrary points tangent with the deformation line with angles $\alpha$ and $\beta$. The deformation along the Z axis modifies the overlap of electron-hole wave functions, this results in observable photo-luminescence variations, see text for details.}
\label{fg:wurz}\end{center}\end{figure}

In this letter we present experimental data and a wave model interpretation that suggests that the discussed optical fluctuations are a general phenomena caused by the interference of thermal vibrational waves that spread along the Quantum Well. Vibration alters the overlap integral, strongly displaced by piezoelectric fields in \ingan materials, and this affects the recombination rate inducing photoluminescence (PL) blinking. This idea is maybe difficult to grasp because thermally induced vibrations are supposed to be incoherent and of very high frequency, whereas PL blinking looks like a slow phenomenon that shows some sort of regularity.\par However, in confined regions, atomic composition and mechanical properties are indeed homogeneous, giving rise to THz range vibrations with limited frequency spread and a certain degree of coherence. In correspondence of dislocations or impurities, mechanical properties of adjacent domains differ slightly, and vibrational interferences are generated. These interferences, simply called {\it beats}, have a broad frequency distribution centred around the frequency difference of the interacting oscillations. The lower frequency tail of this interference pattern is what it is observable as photoluminescence blinking. \par We could confirm the thermal nature of the phenomenon with low temperature investigations. Moreover, using a simple far field correlation methodology we could observe the propagation and spread of PL {\it beats} along the quantum well area up to distances of several tens of microns.

\section{Theoretical insights}

In a Quantum Well structured semiconductor as \ingan, the photoluminescence intensity is regulated by the matrix element that mediate the electron-density and hole-density wave function in the vicinity of the well \cite{chua:1996,chua:1996a}. The spatial overlap of these two function is proportional to the recombination rate, and the latter results in more intense or weaker observable photoluminescence. The difference of inter-atomic distances (lattice mismatch) between the doped and non-doped regions causes piezo-electric stresses that distort the energy bands. This is resulting in diminished wave functions overlap and lower recombination rates due to quantum confined stark effect (QCSE)\cite{rice:2004,das:2011}. \par 
The Wurtzite InGaN quatum well is treated as a two-dimensional system subject to a stress along the direction $xz$; because of spontaneous thermal vibration\cite{jian:2008}, this stress is subject to very small oscillations that alters energy band structure (see the sketch in fig. \ref{fg:wurz}).  

The tension exerted on the line of adjacent atoms is indicated by $T_o$, $ x $ is a direction in the plane of the Quantum Well and $ z $ is the crystal growth vertical axis. We are interested only in displacements along $z$ that are those that influence the matrix elements relative to the recombination rate. We derive the expression for the forces projected over the $z$ axis \begin{equation} \rho dx \ddot{Z}=T_o sin(\alpha)-T_o sin(\beta) \end{equation} where $dx$ is a small displacement along the well, $Z$ is the vertical displacement about the rest position and $ \rho $ is a parameter proportional to the local linear mass density. This approach leads to the general differential expression of a wave oscillating along the crystal growth axis. 

\begin{equation}
 \frac{d^2Z(x,t)}{dt^2}=\frac{T_o}{\rho} \frac{ d^2Z(x,t)}{dx^2}-\frac{\beta}{\rho} \frac{dZ(x,t)}{dt}-\frac{\gamma}{\rho} Z(x,t)
\end{equation}

For generality frictional ($\beta$) and restoring ($\gamma$) parameters are also included. In this classical approximation, we can say that influence of far away atoms is dumped by friction, hence the local solution is a wave of the form $sin(\omega t - kx)$, where $\omega$ is related to $T_o$ and $\rho$ by $\frac{T_o}{\rho}\approx (\frac{\omega}{k})$. 
\par
Elastic mechanical oscillations can interact quasi-coherently in the confined scales, with the term {\it quasi-coherent} we mean that the waves have phase relation in confined ranges, but this relation is lost over longer distances.

Since the phenomenon is driven by spontaneous forces, frequencies are in the order of 1000 cm$^{-1}$ (about 10THz) as shown in Raman studies and related literature \cite{gotz:1996,degu:1999,mich:2004,kuri:2001}. Nevertheless, the observed blinking has components at few seconds range\cite{mich:2006a}. One may ask: how it is possible that the observed slow optical oscillations are induced by such higher frequencies? To answer this, we have to consider that vibrations are localized in a real, non-ideal, crystal system with different Indium concentration, dislocations and impurities that alter the overall vibrational modes. 

In a classic interpretation, this implies that the frequency of oscillation $\omega$, is in reality a spectrum of distributed values that depends on coordinates on the crystal plane $\omega(x,y)$. The mechanical interference of different distributed frequencies results in a wide band of lower frequencies beatings. The beating will be centred aroung $\delta \omega=\omega_1 - \omega_2$, where $ \omega_1 $ and $ \omega_2 $ are the center frequencies of the vibration spectrum relative to two putative adjacent domains. Since mass distribution changes are subtle, $\delta \omega$ can have tails extremely small compared to the original pulsation $\omega$.
\par
We observe optical fluctuations of about  10$\%$~20$\%$ at maximum, experimentation on \ingan devices shows that the quantum well thickness dependence on the intrinsic radiative lifetime is of about 1 order of magnitude per nanometer \cite{berk:1999,bai:2000}. Supposing that the PL oscillation are solely due to this phenomenon, then the greatest spatial displacement along the growth direction is in the range of 0.1 nanometers, a value compatible with the GaN crystal structure.    
\section*{Thermal behaviour}
To prove the mechanical nature of the blinking we study its dependence with temperature. We assume a Boltzmann statistic for the thermal oscillations and we want to verify if the observed photo-luminescence follows the same statistic with temperature. \par
\begin{figure}[!h]
\begin{center}\leavevmode
\includegraphics[width=8.5cm]{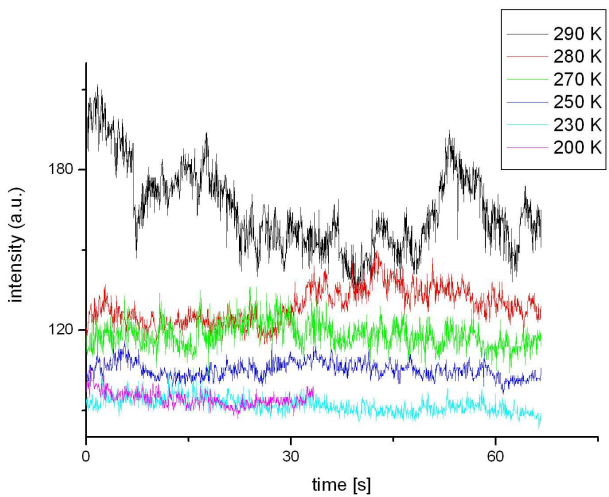}
\caption{The blinking behaviour of a single blinking point at different temperatures.The blinking become less prominent at lower temperature, with a trend compatible with the thermal vibration beating wave hypothesis.}
\label{fg:comparison1}\end{center}\end{figure}

 In a solid we use as potential energy the expression $P.E.=-\frac{1}{2}\frac{YS}{L_o}x^2$, where $Y$ is the Young modulus, $S$ the surface involved in the oscillation, $L_0$ its rest position and $x$ the displacement from it. Assuming a Boltzmann distribution $n=n_0e^{-P.E./kT}$, fixing a displacement one tenth of the maximum calculated above, $x=0.01$ nanometer and using InGaN Young modulus of 350 GPa\cite{jian:2006}, we can derive the potential dependence in a range of temperatures used in the tests (table \ref{tb:BoltzSim_b} and figure \ref{fg:comparison1}).

\begin{table}[h]
\begin{center}
\begin{tabular}{|l|l|r|l|}
\hline
T  & $p_{mn}$ & $p_{mx}$ & $~\delta$ \\
($^\circ$ K) & (a.u.) & (a.u.) & (a.u.) \\
\hline
200 & 1.99 & 1.54 & 0.44 \\
\hline
230 & 4.49 & 3.6 & 0.88 \\
\hline
250 & 6.92 & 5.65 & 1.27 \\
\hline
270 & 10.0 & 8.29 & 1.71 \\
\hline
280 & 11.7 & 9.83 & 1.95 \\
\hline
290 & 13.7 & 11.5 & 2.20 \\
\hline
\end{tabular}
\caption{Table of the predicted thermal behaviour of the system at a set of temperatures according to a Boltzmann statistic. The higher and lower probability values $p_{mx}$, $p_{mn}$ represent a band of $\pm2\%$ in temperature and $\delta$ is the spread. The increased intensity and the spread of luminosity at higher temperatures is expected and confirmed experimentally (fig \ref{fg:comparison1}, \ref{fg:VH}). The elastic potential is calculated using the Young modulus $Y=350$GPa, for simplicity $S=1nm^2$,$L_0=1nm$ and the displacement from the rest position as $x=0.01$ nm, see text for details. }
\label{tb:BoltzSim_b}\end{center}
\vspace{-0.6cm}
\end{table}

\begin{figure}[!h]
\begin{center}\leavevmode
\includegraphics[width=6.5cm]{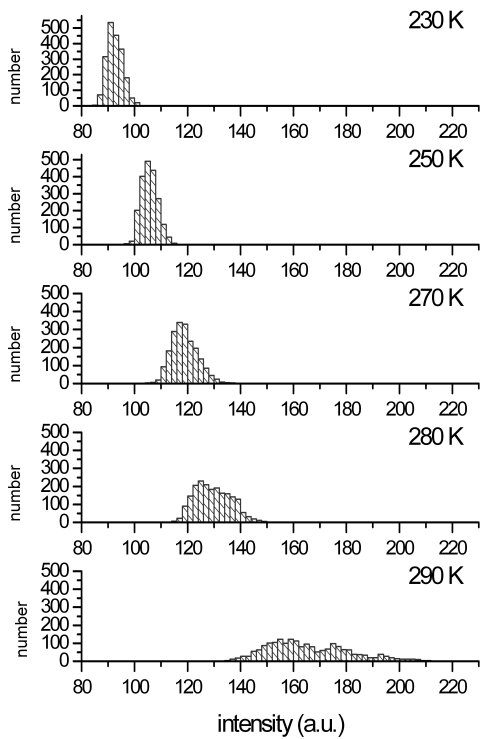}
\caption{Using each temperature of the data in fig. \ref{fg:comparison1}, five histograms are plotted. Increased intensity, and the spread of luminosity at higher temperatures is confirmed in accordance to a vibrational Boltzmann statistic. }
\label{fg:VH}\end{center}\end{figure}

This expected behaviour was confirmed very straightforward tests. A vacuum low temperature chamber was used to observe the blinking at different temperatures ranging from $290$ down to $200$ Kelvin. For lower temperatures not only the average luminosity slightly reduces, but also the variations become less prominent (less blinking) accordingly to thermal model. In figure \ref{fg:comparison1} the raw data are plotted against time, Figure \ref{fg:VH} shows the luminosity distribution histograms at various temperatures. The spread and reduced luminosity of the blinking points predicted is evident. 

\section*{Observation of propagating waves}

\begin{figure}[!h]
\begin{center}\leavevmode
\includegraphics[width=8.5cm]{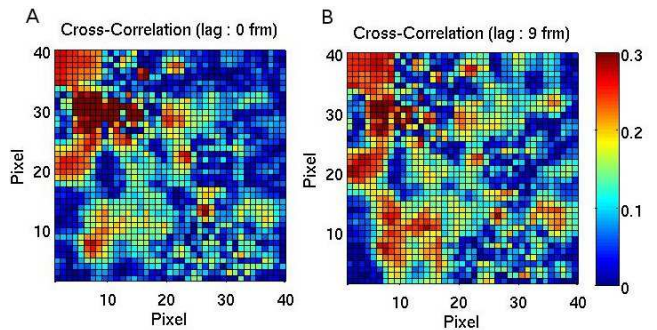}
\caption{The cross-correlation coefficient map to the point located in (10,30) for two different time lags. The blinking behaviour of this single point is correlated with the rest of the map at different times (panel A simultaneously, panel B with a delay equal to 9 frames, about 270 msec). The correlation coefficient decreases in the origin and increases and expands at greater distance with time, suggesting that the "beating" wave is propagating. Correlation values higher than 0.3 have same color for clearer visualization of the expansion of the beating wave. Every pixel is about 1 micrometer.}
\label{fg:cross}\end{center}\end{figure}

We were able to directly detect the horizontal propagation of these interference waves alongside the quantum well. A CCD camera was mounted on a Olympus BX51-W1 microscope to record the photoluminescence on a wide area of the sample. Firstly, we choose a dominant blinking "target" point in the recorded area, then we analysed the time-correlation of every other pixel to this point. The analysis was done in a time span of 120 seconds. In figure \ref{fg:cross} we show the correlation coefficient map referred to a blinking point located at (10,30). The map in panel A show that regions adjacent to the target have an optical dynamics that is correlated to that point, whereas far away points have a more independent dynamics. This behaviour is expected. 

However, when we increase the time lag in the correlation analysis, we observe that as the correlation diminishes in the vicinity of the target as expected, also regions relatively distant from the center of blinking appear to increase in correlation. If we proceed to increase the lag shift, areas of increased correlation expand further, see area around (10,10) in fig. \ref{fg:cross} panel B. This behaviour reveals how the expansion of mechanical beats located on the target are diffusing away from it. The structure and granularity of the correlation coefficient map, shows changing mechanical and mechanical properties along the lattices. 
\par
\section*{Conclusions}
We have realized a simple model and experimental tests that suggest that the optical instabilities in InGaN quantum wells are caused by mechanical beats of thermal vibrations associated with the quantum well lattice. These beats have been observed experimentally through monitoring with CCD camera the photoluminescence and making a time resolved correlation analysis. Furthermore the thermal dependence of the phenomena is compared with a Boltzmann distribution of an harmonic oscillating lattice, resulting in good match and agreement with the original hypothesis.
 \par
We believe that the understanding of the thermal nature of the optical blinking in these devices is important not only for the understanding of the fundamental phenomena involved in the emission of InGaN materials, but also it can be the basis for a new general methodology of analysis of InGaN subtle local mechanical properties, revealing with a very simple and pure optical method, extremely small variation of doping concentration and possibly other fine compositional or structural perturbation in the crystal structure. 

%% Put the bibliography here, most people will use BiBTeX in
%% which case the environment below should be replaced with
%% the \bibliography{} command.


%\bibliographystyle{apsrev4-1}
%\bibliography{biblDatabase_all_20110615}


%% Here is the endmatter stuff: Supplementary Info, etc.
%% Use \item's to separate, default label is "Acknowledgements"

%\begin{footnote}
% \item[Competing Interests] The authors declare that they have no
%competing financial interests.
% \item[Correspondence] Correspondence and requests for materials
%should be addressed to R. M.~(email: ruggero@yokohama-cu.ac.jp).
%\end{footnote}\begin{affiliations}

%%
%% TABLES
%%
%% If there are any tables, put them here.
%%

\end{document}